\documentclass[aps,pre,twocolumn,groupedaddress,showpacs,floatfix]{revtex4}
\usepackage{amsmath}
\usepackage{amssymb}
\usepackage{graphicx}

\begin{document}

\title{Strength Distribution in Gradient Networks}

\author{Luciano da Fontoura Costa} 
\affiliation{Institute of Physics of S\~ao Carlos. 
University of S\~ ao Paulo, S\~{a}o Carlos,
SP, PO Box 369, 13560-970,
phone +55 162 3373 9858,FAX +55 162 3371
3616, Brazil, luciano@if.sc.usp.br}

\date{25th September 2004}

\begin{abstract}   

This article describes a gradient complex network model whose weights
are proportional to the difference between uniformly distributed
``fitness'' values assigned to the nodes.  It is shown analytically
and experimentally that the strength (i.e. the weighted node degree)
density of such a network model can be well approximated by a power
law with $\gamma \approx 0.35$.  Possible implications for neuronal
networks topology and dynamics are also discussed.

\end{abstract}


\maketitle

\section{Introduction}

Great part of the interest focused on complex networks
\cite{Albert_Barab:2002, Newman:2003, Dorog_Mendes:2002} stems from
\emph{scale free} or \emph{power law} distributions of respective
topological measurements, such as the node degree.  At the same time,
weighted complex networks have attracted growing interest because of
their relevance as models of several natural phenomena, with special
attention to systems used for distribution/collection of materials or
information.

There are two main ways to approach the degrees of weighted networks:
(i) by thresholding the weights and using the traditional node degree
\cite{Albert_Barab:2002}; and (ii) by adding the weights of the edges
attached to each node, yielding the respective node \emph{strength}
\cite{Barthelemy_etal:2004}.  Related recent works include the
identification of strength power law in word association networks
\cite{Costa_what:2003, Costa_hier:2004}, studies of scientific
collaborations and air-transportation networks
\cite{Barrat_etal:2003}, the analysis of amino acid sequences in terms
of weighted networks \cite{Costa_hier:2004}, the investigation of
weighted networks defined by dynamical coupling between topology and
weights \cite{Barthelemy_etal:2004}, analytical characterization of
thresholded networks \cite{Masuda_etal:2004}, as well as the
characterization of motifs \cite{Onnela_etal:2004} and shortest paths
in weighted networks \cite{Noh_Rieger:2002}.

The current article describes a gradient oriented network whose nodes
have a respective ``fitness'' value varying uniformly between 0 and 1
with resolution $\Delta_r$, and every node is connected to all other
nodes with higher fitness through an edge with weight proportional to
the difference between the nodes fitness.  It is shown both
analytically and experimentally that the strength densities obtained
for gradient networks can be well approximated by a power law with
parameter $\gamma \approx 0.35$.

We start by describing the gradient network model and follows by
calculating its respective strength distribution, illustrating
experimentally obtained dilog curves, and discussing possible
implications for neuronal networks.

\section{Gradient Networks}

Let the initial state of the directed network $\Gamma$ contain $N$
nodes such that each node $i$ is potentially connected to every other
node $j$, therefore implying a maximum of $N^2-N$ edges.  A
``fitness'' value $v(i)=\lfloor \alpha N_r \rfloor /N_r$ --- where
$\alpha$ is uniformly distributed in the interval $[0, 1)$ and $N_r$
is a positive integer parameter expressing the total of possible
different values $v(i)$ --- is randomly assigned to each node $i$ of
the network.  Consequently, the values $v(i)$ are equally spaced,
i.e. $v(i) = q \Delta_r$ for some $q \in \{0, 1, \ldots, N_r-1\}$,
$\Delta_r = 1/N_r$ and $0 \leq v(i) < 1$. For $N > N_r$, the expected
number of nodes with each specific value $v(i)$ can be estimated as $n
= N \Delta_r$.  The weight of the edge connecting node $i$ to node
$j$, namely $w(i,j)$, with $i \neq j$, is defined as in
Equation~\ref{eq:wgh}.  It follows that $0 \leq w(i,j) < 1$.  Each
weight $w(i,j)$ is henceforth represented as the entry $W(j,i)$ of the
corresponding weight matrix $W$.

\begin{equation}  \label{eq:wgh}
   w(i,j) = \left\{ \begin{array}{ll}
             (v(i) - v(j)) \Delta_r  &   if \,  v(i) > v(j)\\
              0           &   otherwise 
                    \end{array}
            \right.  
\end{equation}

\begin{figure}
 \begin{center} 
   \makebox[4cm][c] { \includegraphics[scale=0.4,angle=-90]{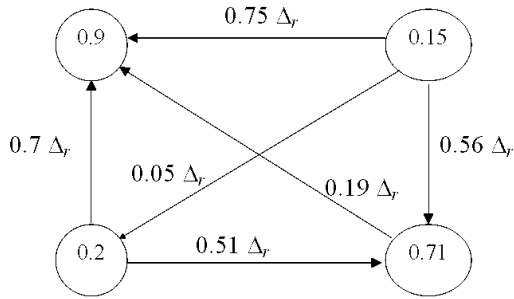} }

   \caption{A simple network obtained by applying the procedure
   described in the text.  Observe that each node receives connections
   from nodes with smaller fitness values, while the respective
   weights are proportional to the fitness
   differences.~\label{fig:ex1}}
\end{center}
\end{figure}

Figure~\ref{fig:ex1} illustrates a simple gradient network obtained by
the above described procedure.  Observe that the weights can be
understood as being proportional to the gradient of $v(i)$ while
moving from any of the neighbors $j$ of $i$, such that $v(j) < v(i)$,
to node $i$, hence the name \emph{gradient networks}.

\section{Analytical Derivation of the Instrength Densities}

The \emph{node instrength} of a node $i$ belonging to a directed
weighted network with weight matrix $W$ is defined as corresponding to
the sum of the weights of the incoming edges, as expressed in
Equation~\ref{eq:indeg}.  Considering a node $i$ with specific
``fitness'' $v(i) = q \Delta_r$, we expect to find, in the average, $n
q$ network nodes with fitness smaller than $v(i)$, namely the nodes
having ``fitness'' $v(i) = j \Delta_r$ with $j = 0, 1, \ldots , q-1$
and respective weights $(q-j)\Delta_r^2 = p \Delta_r^2$, where $p =
q-j = 1, 2, \ldots, q$.  The instrength of node $i$ such that $v(i) =
q \Delta_r$ can thus be estimated as in Equation~\ref{eq:indegi}.

\begin{eqnarray}
   k(q) = \sum_{j=1}^N W(i,j) \label{eq:indeg} \\ 
   k(q) \approx n \sum_{p=1}^{q}  p \Delta_r^2 = 
      \frac{Nq(q+1)\Delta_r^3}{2} \label{eq:indegi}
\end{eqnarray}

By making $c = 0.5 n \Delta_r^2 = 0.5 N \Delta_r^3$, we have

\begin{equation}
   k(q) \approx c (q^2 + q) \label{eq:indegi2}
\end{equation}

The maximum degree is therefore obtained for $v(i) = (N_r-1)\Delta_r$,
yielding $k_{max} \approx 0.5 n$. The fact that we know from the
network construction that a different instrength value is obtained for
each integer instance $q$, allied to the fact that in the average a
total of $n$ nodes will have $v(i)=q \Delta_r$, imply that the
continuous density (over the free variable $x$) of instrengths is
given as in Equation~\ref{eq:pk}.

\begin{equation}
  p(x) = n \sum_{q=0}^{N_r-1} \delta(x-k(q))  \label{eq:pk}
\end{equation}

Let us now consider the estimation of the indstrength distribution by
using the traditional binning scheme given by Equation~\ref{eq:binn},
which maps each indegree $k$ into the respective bin of length
$\Delta$ indexed by $K$.  Observe that symmetric rounding can also be
considered by exchanging the \emph{floor} function by the \emph{round}
function.

\begin{equation}
  K = \lfloor k / \Delta  \rfloor   \label{eq:binn}
\end{equation}

Such a binning procedure can be understood as partitioning the
$x$-axis with interval size $\Delta$.  By approximating the values of
$k$ given in Equation~\ref{eq:indegi2} as $k \approx cq^2$, we have
that $q = \sqrt{k/c}$ and the number of cases expected in each bin ---
yielding the histogram $h(K)$ --- can be estimated as

\begin{equation}
  h(K) \approx n \left( \sqrt{(K+1) \Delta / c} - \sqrt{K \Delta / c} \right)
\end{equation}

By series expansion around $s=0$ it follows that

\begin{equation}
  log(h(s)) \approx \alpha - \gamma s + \beta s^2+ O(s^3) \label{eq:hs} 
\end{equation}

where $\gamma = -0.25 (\sqrt{2}-2)/(\sqrt{2}-1) \approx
0.354$, $\beta = 1/32 (4-3\sqrt{2})/(\sqrt{2}-1)^2 \approx -0.044$
and $\alpha$ is a function of $N$, $\Delta$ and $\Delta_r$. Since the
coefficient of the term in $s^2$ is about 10 times smaller than the
terms in $s$, the dilog curve of the instrength distribution estimated
from the respective histogram for $\Delta \gg \Delta_r$ can be well
approximated by a straight line with slope $-\gamma$.

\section{Discussion}

\begin{figure*}
 \begin{center} 
      \includegraphics[scale=0.7,angle=-90]{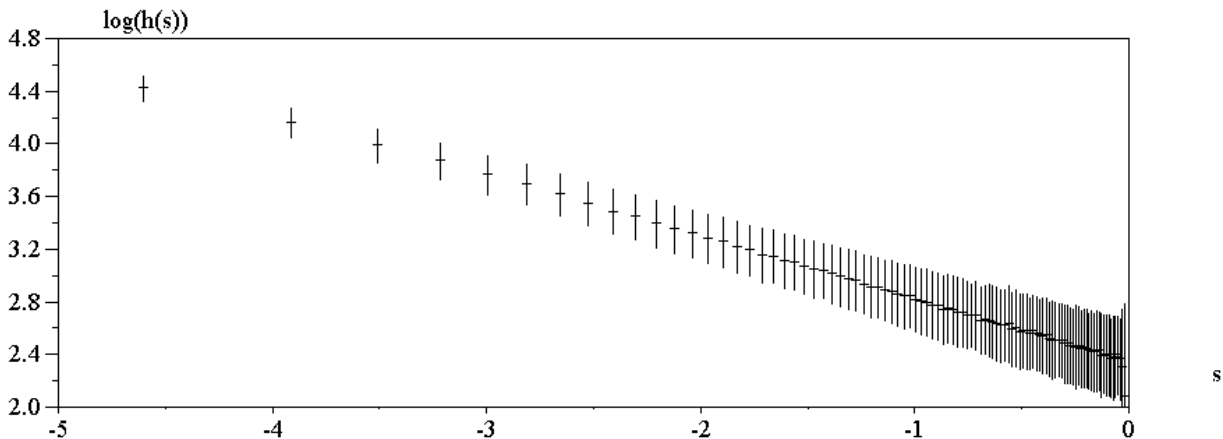} \\

   \caption{Dilog diagram (average $\pm$ standard deviation of
   instrength density (non-cumulative) estimated from the respective
   histogram by using $N=2000$, $\Delta_r=0.001$ and $\Delta = 0.01$.
   An almost straight line segment curve is obtained with slope
   $-0.41$.~\label{fig:bins}}
\end{center}
\end{figure*}

Figure~\ref{fig:bins} shows the dilog plot of instrength density
obtained after 1000 simulations of a gradient network assuming
$N=2000$, $D_r=0.001$ and $\Delta = 0.01$.  As expected, the resulting
curve closely resembles a straight line with estimated (linear
regression) absolute slope value equal to $0.41 \approx \gamma$.  

We have described a simple gradient network model and shown that its
node instrength distribution is close to a power law with parameter
$\gamma \approx 0.35$.  This result implies that although several
nodes of a gradient network have low strength values, there are hubs
characterized by varying large instrength values.  Because the
instrength of a node can be understood as the total weighted influence
it receives from the adjacent nodes, it follows that in case the node
state is directly proportional to its instrength, the dynamics of such
networks will be characterized by a near power law with similar
parameters as those of the instrength density.  Such an interpretation
is particularly interesting from the perspective of integrating
neuronal network and complex network research
(e.g. \cite{Stauffer:2003}, where each neuron is represented by a
node and the synaptic connections by directed edges whose weights
reflect the respective synaptic strengths.  The ``fitness'' values in
such a case could be related to gradient of neurotrophic growth
factors or depolarization bias facilitating the action potential
\cite{Kandel:1995,Freeman:2001}.

It would be interesting to investigate the strength distribution for
``fitness'' statistical models other than that presently considered
uniform density.

\begin{acknowledgments}
The author is grateful to FAPESP (process 99/12765-2), CNPq
(308231/03-1) and the Human Frontier Science Program for financial
support.
\end{acknowledgments}

\bibliography{powerr}

\begin{thebibliography}{13}
\expandafter\ifx\csname natexlab\endcsname\relax\def\natexlab#1{#1}\fi
\expandafter\ifx\csname bibnamefont\endcsname\relax
  \def\bibnamefont#1{#1}\fi
\expandafter\ifx\csname bibfnamefont\endcsname\relax
  \def\bibfnamefont#1{#1}\fi
\expandafter\ifx\csname citenamefont\endcsname\relax
  \def\citenamefont#1{#1}\fi
\expandafter\ifx\csname url\endcsname\relax
  \def\url#1{\texttt{#1}}\fi
\expandafter\ifx\csname urlprefix\endcsname\relax\def\urlprefix{URL }\fi
\providecommand{\bibinfo}[2]{#2}
\providecommand{\eprint}[2][]{\url{#2}}

\bibitem[{\citenamefont{Albert and Barab\'asi}(2002)}]{Albert_Barab:2002}
\bibinfo{author}{\bibfnamefont{R.}~\bibnamefont{Albert}} \bibnamefont{and}
  \bibinfo{author}{\bibfnamefont{A.~L.} \bibnamefont{Barab\'asi}},
  \bibinfo{journal}{Rev. Mod. Phys.} \textbf{\bibinfo{volume}{74}},
  \bibinfo{pages}{47} (\bibinfo{year}{2002}), \bibinfo{note}{cond-mat/0106096}.

\bibitem[{\citenamefont{Newman}(2003)}]{Newman:2003}
\bibinfo{author}{\bibfnamefont{M.~E.~J.} \bibnamefont{Newman}},
  \bibinfo{journal}{SIAM Review} \textbf{\bibinfo{volume}{45}},
  \bibinfo{pages}{167} (\bibinfo{year}{2003}),
  \bibinfo{note}{cond-mat/0303516}.

\bibitem[{\citenamefont{Dorogovtsev and Mendes}(2002)}]{Dorog_Mendes:2002}
\bibinfo{author}{\bibfnamefont{S.~N.} \bibnamefont{Dorogovtsev}}
  \bibnamefont{and} \bibinfo{author}{\bibfnamefont{J.~F.~F.}
  \bibnamefont{Mendes}}, \bibinfo{journal}{Advances in Physics}
  \textbf{\bibinfo{volume}{51}}, \bibinfo{pages}{1079} (\bibinfo{year}{2002}),
  \bibinfo{note}{cond-mat/0106144}.

\bibitem[{\citenamefont{Barth\'elemy et~al.}(2004)\citenamefont{Barth\'elemy,
  Barrat, Pastor-Satorras, and Vespignani}}]{Barthelemy_etal:2004}
\bibinfo{author}{\bibfnamefont{M.}~\bibnamefont{Barth\'elemy}},
  \bibinfo{author}{\bibfnamefont{A.}~\bibnamefont{Barrat}},
  \bibinfo{author}{\bibfnamefont{R.}~\bibnamefont{Pastor-Satorras}},
  \bibnamefont{and}
  \bibinfo{author}{\bibfnamefont{A.}~\bibnamefont{Vespignani}}
  (\bibinfo{year}{2004}), \bibinfo{note}{cond-mat/0408566}.

\bibitem[{\citenamefont{da~F.~Costa}(2004{\natexlab{a}})}]{Costa_what:2003}
\bibinfo{author}{\bibfnamefont{L.}~\bibnamefont{da~F.~Costa}},
  \bibinfo{journal}{Intl. J. Mod. Phys. C} \textbf{\bibinfo{volume}{15}},
  \bibinfo{pages}{371} (\bibinfo{year}{2004}{\natexlab{a}}).

\bibitem[{\citenamefont{da~F.~Costa}(2004{\natexlab{b}})}]{Costa_hier:2004}
\bibinfo{author}{\bibfnamefont{L.}~\bibnamefont{da~F.~Costa}},
  \bibinfo{journal}{Phys. Rev. Lett.} \textbf{\bibinfo{volume}{93}},
  \bibinfo{pages}{098702} (\bibinfo{year}{2004}{\natexlab{b}}).

\bibitem[{\citenamefont{Barrat et~al.}(2003)\citenamefont{Barrat, Barth\'elemy,
  Pastor-Satorras, and Vespignani}}]{Barrat_etal:2003}
\bibinfo{author}{\bibfnamefont{A.}~\bibnamefont{Barrat}},
  \bibinfo{author}{\bibfnamefont{M.}~\bibnamefont{Barth\'elemy}},
  \bibinfo{author}{\bibfnamefont{R.}~\bibnamefont{Pastor-Satorras}},
  \bibnamefont{and}
  \bibinfo{author}{\bibfnamefont{A.}~\bibnamefont{Vespignani}}
  (\bibinfo{year}{2003}), \bibinfo{note}{cond-mat/0311416}.

\bibitem[{\citenamefont{Masuda et~al.}(2004)\citenamefont{Masuda, Miwa, and
  Konno}}]{Masuda_etal:2004}
\bibinfo{author}{\bibfnamefont{N.}~\bibnamefont{Masuda}},
  \bibinfo{author}{\bibfnamefont{H.}~\bibnamefont{Miwa}}, \bibnamefont{and}
  \bibinfo{author}{\bibfnamefont{N.}~\bibnamefont{Konno}}
  (\bibinfo{year}{2004}), \bibinfo{note}{cond-mat/0403524}.

\bibitem[{\citenamefont{Onnela et~al.}(2004)\citenamefont{Onnela, Saramaki,
  K\'ertez, and Kaski}}]{Onnela_etal:2004}
\bibinfo{author}{\bibfnamefont{J.-K.} \bibnamefont{Onnela}},
  \bibinfo{author}{\bibfnamefont{J.}~\bibnamefont{Saramaki}},
  \bibinfo{author}{\bibfnamefont{J.}~\bibnamefont{K\'ertez}}, \bibnamefont{and}
  \bibinfo{author}{\bibfnamefont{K.}~\bibnamefont{Kaski}}
  (\bibinfo{year}{2004}), \bibinfo{note}{cond-mat/0408629}.

\bibitem[{\citenamefont{Noh and Rieger}(2002)}]{Noh_Rieger:2002}
\bibinfo{author}{\bibfnamefont{J.~D.} \bibnamefont{Noh}} \bibnamefont{and}
  \bibinfo{author}{\bibfnamefont{H.}~\bibnamefont{Rieger}},
  \bibinfo{journal}{Phys. Rev. E} \textbf{\bibinfo{volume}{66}},
  \bibinfo{pages}{066127} (\bibinfo{year}{2002}),
  \bibinfo{note}{cond-mat/0208428}.

\bibitem[{\citenamefont{Stauffer et~al.}(2003)\citenamefont{Stauffer, Aharony,
  da~F.~Costa, and Adler}}]{Stauffer:2003}
\bibinfo{author}{\bibfnamefont{D.}~\bibnamefont{Stauffer}},
  \bibinfo{author}{\bibfnamefont{A.}~\bibnamefont{Aharony}},
  \bibinfo{author}{\bibfnamefont{L.}~\bibnamefont{da~F.~Costa}},
  \bibnamefont{and} \bibinfo{author}{\bibfnamefont{J.}~\bibnamefont{Adler}},
  \bibinfo{journal}{Eur. J. Phys.} \textbf{\bibinfo{volume}{32}},
  \bibinfo{pages}{395} (\bibinfo{year}{2003}).

\bibitem[{\citenamefont{Kandel et~al.}(1995)\citenamefont{Kandel, Schwartz, and
  Jessel}}]{Kandel:1995}
\bibinfo{author}{\bibfnamefont{E.~R.} \bibnamefont{Kandel}},
  \bibinfo{author}{\bibfnamefont{J.~H.} \bibnamefont{Schwartz}},
  \bibnamefont{and} \bibinfo{author}{\bibfnamefont{T.~M.}
  \bibnamefont{Jessel}}, \emph{\bibinfo{title}{Essentials of neural science and
  behavior}} (\bibinfo{publisher}{Appleton and Lange},
  \bibinfo{address}{Englewood Cliffs}, \bibinfo{year}{1995}).

\bibitem[{\citenamefont{Freeman}(2001)}]{Freeman:2001}
\bibinfo{author}{\bibfnamefont{W.~J.} \bibnamefont{Freeman}},
  \emph{\bibinfo{title}{How brains make up their minds}}
  (\bibinfo{publisher}{Columbia University Press}, \bibinfo{address}{New York},
  \bibinfo{year}{2001}).

\end{thebibliography}

\end{document}